\documentclass[epj]{svjour}

\usepackage{dsfont}
\usepackage{graphicx}
\usepackage{epstopdf}  
\usepackage{slashed}
\usepackage{amsmath}
\usepackage{float}
\usepackage{tabu}
\usepackage{hyperref}
\allowdisplaybreaks

\begin{document}
\title{The relativistic chiral Lagrangian for decuplet and octet baryons at next-to-leading order}
\author{M\aa ns Holmberg\inst{1} \and Stefan Leupold\inst{1} 
}                     
%
\authorrunning{M.\ Holmberg and S.\ Leupold}
\titlerunning{Chiral Lagrangian for baryons at NLO}
\institute{Institutionen f\"or fysik och astronomi, Uppsala Universitet, Box
516, S-75120 Uppsala, Sweden}
%
\date{\today}
%
\abstract{A complete and minimal relativistic Lagrangian is constructed at next-to-leading order for SU(3) chiral perturbation 
theory in the presence of baryon octet and baryon decuplet states. The Lagrangian has 13 terms for the pure decuplet sector, 
6 terms for the transition sector from baryon octet to decuplet and (as already known from the literature) 16 terms for the pure 
octet sector. The minimal field content of 25 of these terms is meson-baryon four-point interactions.
3 terms give rise to the mass splitting for baryon octet and decuplet states, respectively. 2 terms give rise to overall mass 
shifts. 4 terms provide 
anomalous magnetic moments and a decuplet-to-octet magnetic transition moment. 1 term leads to an 
axial vector transition moment. 
It is shown that meson-baryon three-point coupling constants come in at leading order whereas no additional one 
appears in the minimal Lagrangian at next-to-leading order. Those low-energy constants that give rise to mass splitting 
and magnetic moments, respectively, are determined. Predictions are provided for radiative decays of decuplet to octet baryons.
} 
\maketitle

\section{Introduction and Summary}
\label{sec:intro}

One of the research challenges within the framework of the standard model of particle physics \cite{pesschr} is to understand
the non-perturbative confinement regime of quantum chromodynamics (QCD). 
At (very) low energies chiral perturbation theory 
($\chi$PT) \cite{Weinberg:1978kz,Gasser:1983yg,Gasser:1984gg,Scherer:2002tk,Scherer:2012xha} 
provides a systematically improvable effective field theory to describe and predict QCD phenomena. 
Various versions of $\chi$PT have been 
developed for the lightest two and three quark flavors and for the purely mesonic sector as well as for the sector of the 
lowest lying baryons. In the beginning, a non-relativistic scheme has been used in the baryon sector, but more recently 
relativistic formulations have gained increasing importance; see, e.g., \cite{Scherer:2012xha} for a recent review.

The perturbative expansion 
is carried out in powers of (three-)momenta and masses of the Goldstone bosons. 
Of course, expansion parameters must be 
dimensionless. The ``hard'' scales in the denominator that compensate for the dimensionful ``soft'' momenta and masses 
in the numerator are provided by two distinct QCD properties \cite{Aydemir:2012nz}. On the one hand, it is the scale of
chiral symmetry breaking where loops become as important as tree-level diagrams. On the other hand, it is the scale where 
not considered degrees of freedom become active. If momenta become as large as any of the hard scales, the perturbative expansion
breaks down. In the nucleon sector, the spin-3/2 $\Delta$ baryon sets such a hard scale. 
Quantitatively it is the mass difference between
$\Delta$ and nucleon that comes into play. In practice, this mass difference is as low as $300\,$MeV \cite{pdg}.

If one wants to extend the range of applicability of the effective-field-theory framework, it is conceivable to 
include additional higher-lying degrees of freedom --- provided that it is possible to define a new systematic 
expansion scheme. For the two-flavor sector this implies to include the $\Delta$ degrees of 
freedom, for the three-flavor 
sector it is the whole lowest-lying spin-3/2 decuplet \cite{GellMann:1964nj}. Concerning two flavors see, for instance, 
\cite{Hemmert:1997ye,Hacker:2005fh,Pascalutsa:2005nd,Wies:2006rv,Pascalutsa:2006up} and references therein; for 
three flavors see \cite{Lutz:2001yb,Semke:2005sn,Ledwig:2014rfa}.

Indeed, from the perspective of QCD for a large
number of colors $N_c$ \cite{tHooft:1973alw,Witten:1979kh} the nucleon and $\Delta$ 
are degenerate \cite{Jenkins:1991es,Dashen:1993as}.
In the three-flavor symmetric limit this degeneracy extends to the octet and decuplet. This gives further credit to the idea 
of including the decuplet states as active degrees of freedom in $\chi$PT. Also from a practical point of view, there are 
indications that the convergence properties of baryon $\chi$PT can be improved by a relativistic setup and by the inclusion
of the spin-3/2 states \cite{Jenkins:1991es,Lutz:2001yb,Geng:2008mf,Siemens:2016jwj}.

In the present work we will construct a complete relativistic next-to-leading-order (NLO) Lagrangian for the one-baryon 
three-flavor sector including the lowest-lying octet and decuplet states. The actual challenge is to provide a minimal 
Lagrangian, i.e.\ independent interaction terms that do not lead to the very same observables; 
see also the corresponding discussions in \cite{Ebertshauser:2001nj,Bijnens:1999sh,Fettes:1998ud,Frink:2006hx}. 
This task has been carried out up to (including) next-to-next-to-leading order (NNLO) in \cite{Fettes:1998ud} for the 
two-flavor nucleon sector and in \cite{Frink:2006hx} for the three-flavor baryon-octet sector. 
To the best of our knowledge, the corresponding development for the decuplet sector is missing. We start to fill this gap 
by providing a complete and minimal NLO Lagrangian.

As a second more practical motivation for the present work we note that we are entering a time with increasing experimental
activities in the sector of hyperon spectroscopy. Electromagnetic decays of excited hyperons allow to learn about the 
intrinsic structure of baryons; see also \cite{Granados:2017cib} for an introduction to the physics background.
In particular, the Facility for Antiproton and Ion Research (FAIR) will 
provide excellent opportunities to produce and measure strange hadrons and their radiative decay products (photons and 
electron-positron pairs). For the upcoming phase 0 of FAIR it is planned that the two 
collaborations HADES (High Acceptance DiElectron Spectrometer) \cite{Lorenz:2016qyg} 
and $\bar{\mbox{P}}$ANDA (Antiproton ANnihilation at DArmstadt) \cite{Lutz:2009ff} 
will jointly carry out such analyses.\footnote{K.\ Peters and J.\ Stroth, private communications.}
So far the rather elementary radiative decay processes of a decuplet baryon 
to an octet baryon and a real photon have only been observed for some of the decuplet members \cite{pdg}. We hope that 
this situation will be improved in the near future. 
Therefore we regard it as a good time to provide NLO predictions for these decays as a rough guideline of what to expect. 

Let us give a qualitative summary of the results that will be provided in detail in the main part of this work.
Like in all other sectors of $\chi$PT \cite{Ebertshauser:2001nj,Bijnens:1999sh,Fettes:1998ud,Frink:2006hx}
the majority of higher-order interaction terms gives rise to four-point interactions if one restricts them to their minimal 
field content. In particular, at NLO there are 14 meson-baryon four-point interactions that involve the decuplet states.
They add to the already known \cite{Oller:2006yh} 11 four-point interactions between 
mesons and baryon-octet states. In the absence of stable hyperons and mesons, it is difficult to have enough low-energy data 
to determine all these low-energy constants. For an attempt to pin down parameters in the two-flavor sector 
see \cite{Siemens:2014pma}. 

The minimal field content of the other NLO interaction terms leads to mass shifts and mass splitting and to three-point 
interactions involving external vector and axial vector fields. We will determine as many as possible of the corresponding 
low-energy constants. 

But it is also interesting to discuss which interactions do {\em not} appear at NLO. The leading-order (LO)
Lagrangian provides meson-baryon three-point interactions. Interestingly there are {\em no} additional 
meson-baryon three-point interactions at NLO. This statement applies to our three-flavor case and therefore also carries over 
to the two-flavor case. Yet in the literature, one can find such NLO three-point interactions. But it is easy to understand 
why they are redundant. To this end let us look separately on flavor symmetric and flavor-symmetry breaking 
three-point-interaction terms.

Concerning the flavor symmetric three-point interactions one has 
in general as many independent terms as there are possible partial waves. 
Suppose one has found such a set that accommodates all partial waves. 
Any other term must be on-shell equivalent to the terms of this set  because in a three-point interaction there are no free 
kinematical variables, i.e.\ no energy dependence. Offshell differences between various three-point interactions can be 
compensated by appropriate contact interactions; see, e.g., \cite{Pascalutsa:2000kd}. 

Based on parity symmetry we must have odd orbital angular momentum for the three-point interactions of the pseudoscalar mesons 
with positive-parity baryons. 
One has only p-waves for the three-point interactions of the pseudo\-scalar mesons with baryon octets. 
For the three-point interactions of the pseudo\-scalar mesons with one baryon-octet and one baryon-decuplet field, there is 
also only a p-wave. For the three-point interaction of pseudoscalar mesons with baryon decuplets one can have 
a p- or an f-wave. The latter, however, must be of order ${\cal O}(p^3)$ which is only NNLO.

Flavor breaking three-point interactions appear at order ${\cal O}(p^2)$ but, as already pointed out, 
one has non-van\-ishing orbital angular momentum. 
Thus it is at least a p-wave. Therefore the flavor breaking terms appear only at order ${\cal O}(p^3)$, which is NNLO. 

In the main part of this paper, we collect a couple of technical developments of how to rewrite interaction terms. 
In that way, we hope that the present work helps to sort out which interaction terms are manifestly distinct and which are 
redundant. The rest of the paper is organized in the following way. In section \ref{sec:chiPT} we provide basic definitions 
and the (already established) LO Lagrangian of baryon octet plus decuplet $\chi$PT. The main part of the work is carried 
out in section \ref{sec:NLO} where a complete and minimal NLO Lagrangian is established. Sections \ref{sec:detLEC} and 
\ref{sec:predictrad} are devoted to the determination of some of the NLO low-energy constants. Section \ref{sec:predictrad} 
also provides NLO predictions for radiative decays of decuplet states. Appendices are added to point out some technical
aspects.

\section{Chiral perturbation theory and the leading-order Lagrangian}
\label{sec:chiPT}

The LO chiral Lagrangian 
including the spin-3/2 decuplet states 
is given by \cite{Jenkins:1991es,Lutz:2001yb,Semke:2005sn,Pascalutsa:2006up,Ledwig:2014rfa} 
\begin{eqnarray}
  && {\cal L}_{\rm baryon}^{(1)} = {\rm tr}\left(\bar B \, (i \slashed{D} - m_{(8)}) \, B \right)  \nonumber \\ 
  && {}+ \bar T_{abc}^\mu \, ( i \gamma_{\mu\nu\alpha} (D^\alpha T^\nu)^{abc} - \gamma_{\mu\nu} \, m_{(10)} \, (T^\nu)^{abc})
  \nonumber \\ 
  && {}+ \frac{D}{2} \, {\rm tr}(\bar B \, \gamma^\mu \, \gamma_5 \, \{u_\mu,B\}) 
  + \frac{F}{2} \, {\rm tr}(\bar B \, \gamma^\mu \, \gamma_5 \, [u_\mu,B])  \nonumber \\
  && {} + \frac{h_A}{2\sqrt{2}} \, 
  \left(\epsilon^{ade} \, \bar T^\mu_{abc} \, (u_\mu)^b_d \, B^c_e
    + \epsilon_{ade} \, \bar B^e_c \, (u^\mu)^d_b \, T_\mu^{abc} \right) \nonumber \\
  && {} -\frac{H_A}{2} \, \bar T^\mu_{abc} \gamma_\nu \gamma_5 \, (u^\nu)^c_d \; T_\mu^{abd} 
  \label{eq:baryonlagr}
\end{eqnarray}
with tr denoting a flavor trace. 

We have introduced the totally antisymmetrized products of two and three 
gamma matrices\footnote{Throughout this work, when using the phrase ``gamma matrices'' we have the four gamma 
matrices $\gamma^\mu$, $\mu=0,1,2,3$, in mind, {\em not} $\gamma_5$.} \cite{pesschr},
\begin{eqnarray}
  \label{eq:defgammunu}
  \gamma^{\mu\nu} := \frac12 [\gamma^\mu,\gamma^\nu] = -i \sigma^{\mu\nu}
\end{eqnarray}
and 
\begin{eqnarray}
  \label{eq:defgammunual}
  \gamma^{\mu\nu\alpha}&:=& \frac16 
  \left(\gamma^\mu \gamma^\nu \gamma^\alpha + \gamma^\nu \gamma^\alpha \gamma^\mu + \gamma^\alpha \gamma^\mu \gamma^\nu
  \right.  \nonumber \\ && \left. \phantom{m} {}
    - \gamma^\mu \gamma^\alpha \gamma^\nu - \gamma^\alpha \gamma^\nu \gamma^\mu - \gamma^\nu \gamma^\mu \gamma^\alpha \right)
  \nonumber \\ 
  & = & \frac12 \{\gamma^{\mu\nu},\gamma^\alpha\} = +i\epsilon^{\mu\nu\alpha\beta} \gamma_\beta \gamma_5  \,,
\end{eqnarray}
respectively.
Our conventions are: $\gamma_5:=i \gamma^0 \gamma^1 \gamma^2 \gamma^3$ and 
$\epsilon_{0123}=-1$ (the latter in agreement with \cite{pesschr} 
but opposite to \cite{Pascalutsa:2006up,Ledwig:2011cx}). If a formal manipulation program is used to calculate spinor traces and 
Lorentz contractions a good check for the convention for the Levi-Civita symbol is the last relation in (\ref{eq:defgammunual}).

The spin-1/2 octet baryons are collected in ($B^a_b$ is the entry in the $a$th row, $b$th column)
\begin{eqnarray}
  \label{eq:baroct}
  B = \left(
    \begin{array}{ccc}
      \frac{1}{\sqrt{2}}\, \Sigma^0 +\frac{1}{\sqrt{6}}\, \Lambda 
      & \Sigma^+ & p \\
      \Sigma^- & -\frac{1}{\sqrt{2}}\,\Sigma^0+\frac{1}{\sqrt{6}}\, \Lambda
      & n \\
      \Xi^- & \Xi^0 
      & -\frac{2}{\sqrt{6}}\, \Lambda
    \end{array}   
  \right)  \,.
\end{eqnarray}
The decuplet is expressed by a totally symmetric flavor tensor $T^{abc}$ 
with 
\begin{eqnarray}
  && T^{111} = \Delta^{++} , \quad T^{112} = \frac{1}{\sqrt{3}} \, \Delta^+  , \nonumber \\
  && T^{122} = \frac{1}{\sqrt{3}} \, \Delta^0  , \quad T^{222} = \Delta^- ,    \nonumber \\
  && T^{113} = \frac{1}{\sqrt{3}} \, \Sigma^{*+}  , \quad T^{123} = \frac{1}{\sqrt{6}} \, \Sigma^{*0}  , \quad 
  T^{223} = \frac{1}{\sqrt{3}} \, \Sigma^{*-}  ,  \nonumber \\
  && T^{133} = \frac{1}{\sqrt{3}} \, \Xi^{*0} , \quad T^{233} = \frac{1}{\sqrt{3}} \, \Xi^{*-} , \quad 
  T^{333} = \Omega \,. 
  \label{eq:tensorT}
\end{eqnarray}
The Goldstone bosons are encoded in
\begin{eqnarray}
  \Phi &=&  \left(
    \begin{array}{ccc}
      \pi^0 +\frac{1}{\sqrt{3}}\, \eta 
      & \sqrt{2}\, \pi^+ & \sqrt{2} \, K^+ \\
      \sqrt{2}\, \pi^- & -\pi^0+\frac{1}{\sqrt{3}}\, \eta
      & \sqrt{2} \, K^0 \\
      \sqrt{2}\, K^- & \sqrt{2} \, {\bar{K}}^0 
      & -\frac{2}{\sqrt{3}}\, \eta
    \end{array}   
  \right) 
  \,, \nonumber \\
  u^2 & := & U := \exp(i\Phi/F_\pi) \,, \quad u_\mu := i \, u^\dagger \, (\nabla_\mu U) \, u^\dagger = u_\mu^\dagger \,. \phantom{mm}
  \label{eq:gold}
\end{eqnarray}

The fields have the following transformation properties with respect to chiral 
transformations \cite{Jenkins:1991es,Scherer:2012xha}
\begin{eqnarray}
  U \to L \, U \, R^\dagger \,, &&  u \to L \, u \, h^\dagger = h \, u \, R^\dagger  \,, \nonumber \\
  \label{eq:chiraltrafos}  
  u_\mu \to h \, u_\mu \, h^\dagger \,, &&  B \to h \, B \, h^\dagger \,,   \\
  T^{abc}_\mu \to h^a_{d} \, h^b_{e} \, h^c_{f} \, T^{def}_\mu \,,  && 
  \bar T_{abc}^\mu \to (h^\dagger)_a^{d} \, (h^\dagger)_b^{e} \, (h^\dagger)_c^{f} \, \bar T_{def}^\mu \,.  \nonumber 
\end{eqnarray}
In particular, the choice of upper and lower flavor indices is used to indicate that upper indices transform with $h$ 
under flavor transformations while the lower components transform with $h^\dagger$. 

The chirally covariant derivative for a (baryon) octet is defined by
\begin{eqnarray}
  \label{eq:devder}
  D^\mu B := \partial^\mu B + [\Gamma^\mu,B]   \,,
\end{eqnarray}
for a decuplet $T$ by
\begin{eqnarray}
  (D^\mu T)^{abc} &:=& \partial^\mu T^{abc} + (\Gamma^\mu)^a_{a'} T^{a' bc} + (\Gamma^\mu)^b_{b'} T^{a b' c} \nonumber \\
  && {} + (\Gamma^\mu)^c_{c'} T^{a bc'}   \,,
  \label{eq:devderdec}
\end{eqnarray}
for an anti-decuplet by 
\begin{eqnarray}
  (D^\mu \bar T)_{abc} &:=& \partial^\mu \bar T_{abc} - (\Gamma^\mu)_a^{a'} \bar T_{a' bc} - (\Gamma^\mu)_b^{b'} \bar T_{a b' c}  
  \nonumber \\
  && {} - (\Gamma^\mu)_c^{c'} \bar T_{a bc'}   \,,
  \label{eq:devderantidec}
\end{eqnarray}
and for the Goldstone boson fields by
\begin{eqnarray}
  \label{eq:devderU}
  \nabla_\mu U := \partial_\mu U -i(v_\mu + a_\mu) \, U + i U \, (v_\mu - a_\mu)
\end{eqnarray}
with
\begin{eqnarray}
  \Gamma_\mu &:=&  \frac12 \, \left(
    u^\dagger \left( \partial_\mu - i (v_\mu + a_\mu) \right) u \right. \nonumber \\
    && \phantom{m} \left. {}+
    u \left( \partial_\mu - i (v_\mu - a_\mu) \right) u^\dagger
  \right) \,,
  \label{eq:defGammamu}
\end{eqnarray}
where $v$ and $a$ denote external sources. 

In (\ref{eq:baryonlagr}) $m_{(8)}$ ($m_{(10)}$) denotes the mass of the baryon octet (decuplet) in the chiral limit. 
For the octet and for the decuplet the 
flavor breaking terms that appear at NLO are capable of splitting up the baryon masses 
such that they 
are sufficiently close to the physical masses. This will be discussed in section \ref{sec:detLEC}.

Standard values for the coupling constants are $F_\pi=92.4\,$MeV, $D=0.80$, $F=0.46$. For the pion-nucleon coupling constant this 
implies $g_A=F+D =1.26$. 

The value for $h_A$ can be determined from
the partial decay width $\Sigma^* \to \pi \, \Lambda$ or $\Sigma^* \to \pi \, \Sigma$ 
yielding $h_A=2.3\pm 0.1$ \cite{Granados:2017cib}. In the cascade sector the better determined decay width comes from the neutral
cascade \cite{pdg}. This leads to $h_A = 2.00 \pm 0.06$. 
From the $\Delta \to \pi \, N$ decay one finds $h_A = 2.88$ \cite{Granados:2017cib}. Note that the determination of partial
decay widths is easier for narrow states than for broad resonances. Thus the extraction of $h_A$ from hyperon decays is 
preferable. 
One might also look at large-$N_c$ estimates for two or three flavors:
$h_A=3 g_A/\sqrt{2} \approx 2.67$ according to \cite{Pascalutsa:2005nd,Pascalutsa:2006up,Ledwig:2011cx} 
or $h_A = 2\sqrt{2} D \approx 2.26$ according to \cite{Dashen:1993as,Semke:2005sn}. All these values are in the range
$h_A = 2.4 \pm 0.5$. Such a flavor breaking effect of 20\%\ appears completely acceptable for a leading-order calculation of the
decay widths. 

Finally one has to specify $H_A$. In absence of a simple direct observable to pin it down, we take estimates from large-$N_c$ 
considerations: $H_A = \frac95 \, g_A \approx 2.27$ \cite{Pascalutsa:2006up,Ledwig:2011cx} 
or $H_A = 9F -3D \approx 1.74$ \cite{Dashen:1993as,Semke:2005sn}.
We have checked explicitly that the sign of $H_A$ is in 
agreement with \cite{Pascalutsa:2006up,Ledwig:2011cx} and also with \cite{Semke:2005sn}. For quark-model estimates of these
coupling constants see \cite{Buchmann:1999ab,Buchmann:2013fxa}. 

In (\ref{eq:baryonlagr}) we have written down the simplest interaction terms for the decuplet baryons. Their Lorentz and 
spinor structure is given by vector-spinors as suggested in \cite{Rarita:1941mf}. A somewhat unpleasant 
feature of these Rarita-Schwinger fields is the fact that they describe not only spin-3/2 degrees of freedom 
but in addition spin-1/2 modes. The free Lagrangian is chosen such that the spin-1/2 modes are frozen. Only the spin-3/2 modes
constitute propagating degrees of freedom. In other words, the free Lagrangian produces constraint equations together with the
equations of motion. Yet in the presence of interactions the spin-1/2 modes can contribute. Actually, they give rise to additional
contact interactions. There are 
several suggestions in the literature how to deal with this influence of the frozen spin-1/2 modes. 
One possibility is to 
analyze the structure of the constraints and its impact on the coupling constants for the most 
general interaction terms \cite{Hacker:2005fh,Wies:2006rv}. 
Another possibility is to construct the 
interaction terms such that the spin-1/2 modes are projected away \cite{Pascalutsa:1999zz,Pascalutsa:2005nd,Pascalutsa:2006up}.
For instance, for the $H_A$ term in (\ref{eq:baryonlagr}), the rewritten interaction term 
reads \cite{Pascalutsa:2006up,Ledwig:2014rfa} 
\begin{eqnarray}
  && - \frac{H_A}{4 m_{(10)}} \, 
  \left(\bar T^\mu_{abc} \, (D^\nu T^\alpha)^{abd} 
    + (D^\nu \bar T^\alpha)_{abc} \, (T^\mu)^{abd} \right) 
  \nonumber \\ && {} \phantom{m} \times 
  \epsilon_{\mu\nu\alpha\beta} \, (u^\beta)^c_d   \,.
  \label{eq:HA-Pasc}  
\end{eqnarray}
On the other hand, it can be shown that any interaction term that can be constructed at leading order is on-shell 
equivalent to the ones presented in (\ref{eq:baryonlagr}). Any difference can be accounted for by explicit contact terms 
of the NLO Lagrangian \cite{Pascalutsa:2000kd,Granados:2017cib}. Therefore we stick to the simplest LO interaction terms.

\section{The next-to-leading-order Lagrangian}
\label{sec:NLO}

A complete NLO Lagrangian for the baryon-octet sector is given in \cite{Oller:2006yh,Frink:2006hx}.
But a complete NLO Lagrangian including the decuplet baryons has not been constructed so far. 
It can be split up into a transition part 
involving one baryon-decuplet and one baryon-octet field and in a part involving solely the decuplet fields. 
In the next subsection we will construct the complete and minimal NLO Lagrangian for the transition part.
In subsection \ref{sec:decNLO} we will address the pure decuplet sector. In subsection \ref{sec:octetNLO} we comment 
on the construction of the NLO Lagrangian for the baryon-octet sector.

The NLO Lagrangian contains all independent terms of order $p^2$ where $p$ denotes a small momentum or Goldstone boson mass.
The pertinent building blocks of ${\cal O}(p^2)$ are \cite{Ecker:1988te,Fettes:1998ud,Bijnens:1999sh,Ebertshauser:2001nj}
  \begin{eqnarray}
    \label{eq:op2struc}
    \chi_\pm \,, \quad \mbox{and} \quad {\cal O}_2^{\mu\nu} = f_\pm^{\mu\nu}, u^\mu u^\nu , D^\mu u^\nu + D^\nu u^\mu 
  \end{eqnarray}
with $\chi_\pm = u^\dagger \chi u^\dagger \pm u \chi^\dagger u$ and $\chi = 2 B_0 \, (s+ip)$ 
obtained from the scalar source $s$ and the pseudoscalar source $p$. The scalar source contains the quark mass matrix. Formally
$\chi_\pm$ constitute flavor nonets, i.e.\ octet plus singlet.
The low-energy constant $B_0$ is essentially the ratio of
the light-quark condensate and the square of the pion-decay constant; 
see, e.g.\ \cite{Gasser:1983yg,Gasser:1984gg,Scherer:2002tk,Scherer:2012xha}. 

The field strengths are given by
\begin{eqnarray}
  \label{eq:deffpm}
  f_\pm^{\mu\nu} := u \, F_L^{\mu\nu} \, u^\dagger \pm u^\dagger \, F_R^{\mu\nu} \, u
\end{eqnarray}
with
\begin{equation}
  F_{R,L}^{\mu\nu} := \partial^\mu \, (v^\nu \pm a^\nu) - \partial^\nu \, (v^\mu \pm a^\mu) -i \, [v^\mu \pm a^\mu,v^\nu \pm a^\nu]  \,.
  \label{eq:defFRL}
\end{equation}
The field strengths constitute flavor octets. Interactions with electromagnetism 
can be studied by the replacement \cite{Scherer:2002tk}
\begin{eqnarray}
  \label{eq:emvA}
  v_\mu \to e A_\mu \left(
    \begin{array}{rrr}
      \frac23 & 0 & 0  \\[0.5em]
      0 & -\frac13 & 0 \\[0.5em]
      0 & 0 & -\frac13
    \end{array}
    \right)
\end{eqnarray}
with the photon field $A_\mu$ and the proton charge $e$.

In (\ref{eq:op2struc}), only the symmetric combination of derivative and $u^\mu$ appears because the antisymmetric combination
satisfies \cite{Bijnens:1999sh}:
\begin{eqnarray}
  \label{eq:Duf-}
  D^\mu u^\nu - D^\nu u^\mu = f_-^{\mu\nu}   \,.
\end{eqnarray}

Let us add some general remarks: Interaction terms can be rewritten with the use of the equations of motion that emerge 
from the LO Lagrangian; see, e.g., \cite{Fettes:1998ud,Bijnens:1999sh,Ebertshauser:2001nj}. 
One can show that the differences are beyond the order that one considers. 
The LO equations of motion (including constraint equations) read
\begin{eqnarray}
  (i \slashed{D} - m_{(8)}) \, B  & = & {\cal O}(p)   \,, \nonumber  \\
  (i \slashed{D} - m_{(10)}) \, T_\mu & = & {\cal O}(p)   \,, \nonumber  \\
  \gamma^\mu T_\mu & = & {\cal O}(p)   \,, \nonumber  \\
  D^\mu T_\mu & = & {\cal O}(p)   \,. 
  \label{eq:LO-EOMs}  
\end{eqnarray}
Here $p$ denotes a soft momentum. It is important to note that on the respective right-hand side of (\ref{eq:LO-EOMs}) 
there is always a $u^\mu$ (at order $p$).

In general, (chiral) derivatives acting on the baryon fields do not produce small momenta. 
In other words such terms are not suppressed.
However, commutators of such derivatives are suppressed \cite{Fettes:1998ud,Bijnens:1999sh}:
\begin{eqnarray}
  \label{eq:supprDD}
  [D^\mu,D^\nu] \, B = [\Gamma^{\mu\nu},B] 
\end{eqnarray}
with 
\begin{eqnarray}
  \label{eq:supprDD2}
  \Gamma^{\mu\nu} := \frac14 \, [u^\mu,u^\nu] - \frac{i}{2} \, f_+^{\mu\nu} = {\cal O}(p^2)   \,.
\end{eqnarray}
A corresponding relation holds for the decuplet baryons. 
As a consequence one obtains
\begin{eqnarray}
  \label{eq:towardsKG}
  (i \slashed{D} + m)(i \slashed{D} - m) = -D^2-m^2 + {\cal O}(p^2)
\end{eqnarray}
and therefore the LO Klein-Gordon equations read
\begin{eqnarray}
  \label{eq:KGLO}
  (D^2 + m_{(8)}^2) \, B  & = & {\cal O}(p)   \,, \nonumber  \\
  (D^2 + m_{(10)}^2) \, T_\mu & = & {\cal O}(p)  \,.
\end{eqnarray}

\subsection{Transition sector}
\label{sec:transNLO}

For completeness let us first construct the LO Lagrangian for the transition part. The building blocks are
\begin{itemize}
\item one $\bar B$ and one $T_\mu$;
\item one or no $u_\nu$;
\item arbitrary many gamma matrices;
\item arbitrary many chiral derivatives acting on $T_\mu$;
\item a $\gamma_5$ if required by parity symmetry.
\end{itemize}
Note that a Levi-Civita symbol can be traded in for a $\gamma_5$ and a couple of gamma matrices \cite{pesschr}. 
Thus we do not consider it 
separately. Up to total derivatives that do not change the equations of motion, the derivatives acting on $\bar B$ can be 
reshuffled such that they act on $T_\mu$ plus terms of order ${\cal O}(p^2)$. 

If there is no $u_\nu$ then we have only one octet and one decuplet. This cannot be combined to a flavor singlet. Thus one 
needs the $u_\nu$. 
Terms where gamma matrices are contracted with each other can be simplified by using the anticommutation relations of 
gamma matrices. One obtains terms with less many gamma matrices. Terms where derivatives are contracted with gamma matrices
or with themselves can be simplified by the equations of motion (\ref{eq:LO-EOMs}) or the Klein-Gordon equation (\ref{eq:KGLO}).
The same holds for terms where $T_\mu$ is contracted with a gamma matrix or a derivative. Thus $T_\mu$ must be contracted 
with $u^\mu$. This leads to the $h_A$ term of (\ref{eq:baryonlagr}). 

The building blocks for the transition part of the NLO Lagrangian are 
\begin{itemize}
\item one $\bar B$ and one $T_\mu$;
\item one of the ${\cal O}(p^2)$ structures given in (\ref{eq:op2struc});
\item arbitrary many gamma matrices;
\item arbitrary many chiral derivatives acting on $T_\mu$;
\item a $\gamma_5$ if required by parity symmetry.
\end{itemize}
Note that one of the ${\cal O}(p^2)$ structures must appear. Otherwise the case is already covered by our previous considerations.
Up to total derivatives that do not change the equations of motion, the derivatives acting on $\bar B$ can be 
reshuffled such that they act on $T_\mu$ plus terms of order ${\cal O}(p^3)$. 

With the same considerations as before we can rewrite terms with gamma matrices and derivatives contracted among themselves 
or with $T_\mu$. 
Thus $T_\mu$ must be contracted with
a structure from (\ref{eq:op2struc}). This excludes $\chi_\pm$ and leaves us with one of the ${\cal O}_2^{\mu\nu}$ structures. 
One index of ${\cal O}_2^{\mu\nu}$ is contracted with $T_\mu$. The second index must be contracted with 
a gamma matrix or with a hard derivative acting on $T_\mu$. 

We will show now that the latter case can be rewritten into the former \cite{Fettes:1998ud}.
Note that the structure that we will write down next has a yet to be specified flavor structure. For the arguments that we use
this flavor structure is, however, not important. Apart from the flavor structure, the NLO terms with one hard derivative read
\begin{eqnarray}
  && \bar B \, (\gamma_5) \, {\cal O}_2^{\mu\nu} i D_\mu T_\nu 
  = \bar B \, (\gamma_5) \, {\cal O}_2^{\mu\nu} g_{\mu \alpha} \, i D^\alpha T_\nu   \nonumber \\
  && = \bar B \, (\gamma_5) \, {\cal O}_2^{\mu\nu} \frac12 
  (\gamma_\mu \gamma_\alpha + \gamma_\alpha \gamma_\mu) \, i D^\alpha T_\nu  \nonumber \\
  && = \frac12 \bar B \, (\gamma_5) \, {\cal O}_2^{\mu\nu} (\gamma_\mu \, i \slashed{D} + i\slashed{D} \, \gamma_\mu) \, T_\nu  \,.
  \label{eq:groD}
\end{eqnarray}
The appearance or absence of $\gamma_5$ depends on the choice for ${\cal O}_2^{\mu\nu}$. 
It is dictated by parity symmetry, see Appendix \ref{sec:discsym} below. 
For the first term in the sum that appears in the last expression in (\ref{eq:groD}), 
we use the equation of motion from (\ref{eq:LO-EOMs}). Thus the hard derivative is effectively replaced by 
a gamma matrix. For the second term we use integration by parts (in the action) to reshuffle the derivative such that it 
acts on $\bar B$. 
The extra term where the derivative acts on ${\cal O}_2^{\mu\nu}$ is of order ${\cal O}(p^3)$. When the derivative 
acts on $\bar B$ we can use the Dirac equation from (\ref{eq:LO-EOMs}). In case that $\gamma_5$ is present one uses the 
Dirac equation after anticommutation. Again we find that the derivative is 
effectively replaced by a gamma matrix. 

The previous considerations have shown that apart from the flavor structure there are only four types of terms:
\begin{eqnarray}
  && \bar B \gamma_\mu \gamma_5 f_+^{\mu\nu} T_\nu \,, \quad  \bar B \gamma_\mu f_-^{\mu\nu} T_\nu \,, \quad  
  \bar B \gamma_\mu \gamma_5 u^\mu u^\nu T_\nu \,, \nonumber \\  
  && \bar B \gamma_\mu (D^\mu u^\nu + D^\nu u^\mu) T_\nu   \,.
  \label{eq:gen3}
\end{eqnarray}
We note again that the appearance or absence of $\gamma_5$ is dictated by parity symmetry, see Appendix \ref{sec:discsym} below.
We will show now that the last term of (\ref{eq:gen3}) is redundant. 
This term induces (flavor symmetric) three-point couplings between 
a decuplet baryon, an octet baryon and a pseudoscalar meson, just like the LO term $\sim h_A$ from (\ref{eq:baryonlagr})
does. There is only one partial wave for the corresponding decay, a p-wave. Therefore it should be clear that all flavor 
symmetric terms that introduce this three-point coupling must be on-shell equivalent. In other words, in a complete and minimal
NLO Lagrangian the last term of (\ref{eq:gen3}) should not appear.

Using (\ref{eq:Duf-}) we obtain
\begin{eqnarray}
  \label{eq:4th-term}
  \bar B \gamma_\mu (D^\mu u^\nu + D^\nu u^\mu) T_\nu = \bar B \gamma_\mu (2 D^\mu u^\nu - f_-^{\mu\nu} ) T_\nu \,.
\end{eqnarray}
The term $\sim f_-^{\mu\nu}$ appears explicitly in (\ref{eq:gen3}). Thus we can concentrate on 
\begin{equation}
  \label{eq:4th-term2}
  \bar B \gamma_\mu D^\mu u^\nu T_\nu = - \bar B \overleftarrow{\slashed{D}} u^\nu T_\nu - \bar B u^\nu \slashed{D} \, T_\nu  
  + \mbox{total derivative}   \,.
\end{equation}
Using again the equations of motion (\ref{eq:LO-EOMs}) produces terms that resemble the $h_A$ term of the LO 
Lagrangian (\ref{eq:baryonlagr}) and terms that contain $u^\nu u^\mu$. 
In this context, we recall that the right-hand side of (\ref{eq:LO-EOMs}) brings in the 
additional $u^\mu$. 
But terms of the type $\sim u^\nu u^\mu$ appear 
explicitly in (\ref{eq:gen3}) if they contain $\bar B$ and $T_\nu$, i.e.\ if they belong at all to the transition sector. 
Terms of the pure baryon-octet sector $\sim \bar B \ldots B$ or of the pure decuplet sector $\sim \bar T^\mu \ldots T^\nu$ 
are not of our concern for the moment. But also in these sectors, terms with $u^\nu u^\mu$ are accounted for explicitly, as
discussed in the next two subsections.

We have shown that only the first three terms in (\ref{eq:gen3}) contribute to a complete and minimal NLO Lagrangian for 
the octet-to-decuplet transition sector. Finally, we have to pin down all possibilities for the flavor structure.
The first two terms in (\ref{eq:gen3}) involve two flavor octets and one decuplet. 
There is only one way to construct a flavor-symmetry invariant 
interaction Lagrangian, i.e.\ a flavor singlet:
\begin{eqnarray}
  && i \left( \epsilon_{ade} \, \bar B^e_c \, \gamma_\mu (\gamma_5) (f_{\pm}^{\mu\nu})^d_b \, T_\nu^{abc} 
  \right. \nonumber \\ 
  &&\phantom{m} \left. {}
    - \epsilon^{ade} \, (\bar T_\nu)_{abc} \, \gamma_\mu (\gamma_5) (f_\pm^{\mu\nu})_d^b \, B_e^c \right) \,. 
  \label{eq:finalfpm}
\end{eqnarray}
We recall that $f_+$ comes with a $\gamma_5$, while $f_-$ does not.

The construction of the pertinent flavor structures for the third term in (\ref{eq:gen3}) is more complex. We will involve 
some group theory documented in Appendix \ref{sec:sgt}. The result of all our considerations is 
\begin{eqnarray}
  {\cal L}_{8-10}^{(2)} &= &
  i \, c_M \left( \epsilon_{ade} \, \bar B^e_c \, \gamma_\mu \gamma_5 (f_+^{\mu\nu})^d_b \, T_\nu^{abc} \right. \nonumber \\ 
  &&\phantom{mmm} \left. {}
    - \epsilon^{ade} \, (\bar T_\nu)_{abc} \, \gamma_\mu \gamma_5 (f_+^{\mu\nu})_d^b \, B_e^c \right) \nonumber \\
  && {} + i \, c_E \left( \epsilon_{ade} \, \bar B^e_c \, \gamma_\mu (f_-^{\mu\nu})^d_b \, T_\nu^{abc} \right. \nonumber \\
  &&\phantom{mmmm} \left. {}
    -\epsilon^{ade} \, (\bar T_\nu)_{abc} \, \gamma_\mu (f_-^{\mu\nu})_d^b \, B_e^c \right) \nonumber \\
  && {} + c_F \, ({\cal O}_{\mu\nu})^b_a \, ({\cal O}_F^{\mu\nu})^a_b 
  + c_D \, ({\cal O}_{\mu\nu})^b_a \, ({\cal O}_D^{\mu\nu})^a_b   \nonumber \\ &&
  {} + c_{(10)} \, {\cal D}^{abc}_{\mu\nu} \, (\bar{\cal D}_M^{\mu\nu})_{abc}  
  + c_{(27)} \, ({\cal S}_{\mu\nu})^{cd}_{ab} \, ({\cal S}_M^{\mu\nu})^{ab}_{cd} \nonumber \\ && {}+ {\rm h.c.}  
  \label{eq:transNLO}
\end{eqnarray}
where ``h.c.'' denotes the hermitian conjugate for the preceding four terms. The appearance or absence of $i$'s is chosen such
that the coupling constants $c_{\ldots}$ are real if charge conjugation symmetry holds.

\subsection{Decuplet sector}
\label{sec:decNLO}

The procedure resembles the one for the transition sector. But now we have two vector-spinors for the baryon fields. 
For the construction of the pertinent interaction terms, it is useful to know the transformation behavior with respect to 
parity flip and charge conjugation. This information is summarized in Appendix \ref{sec:discsym}. 

First, we discuss briefly the LO structures. If no $u_\mu$ appears 
one obtains the terms of the (chiralized) free Lagrangian. For terms with one $u_\mu$ one might contract it with one baryon 
vector-spinor. But then the other baryon field would be contracted with a derivative or a gamma matrix which leads to subleading
terms on account of the constraint equations in (\ref{eq:LO-EOMs}). Thus the two baryon fields must be contracted with each 
other while $u_\mu$ is contracted with a derivative or a gamma matrix. Parity enforces the appearance of $\gamma_5$. 
However, the combination of $\gamma_5$ and a derivative (acting on a baryon field) is odd with respect to charge conjugation.
Thus we have finally exactly one 
term with one $u_\mu$ in the LO Lagrangian. This is the $H_A$ term that appears in (\ref{eq:baryonlagr}).

Based on the previous considerations the NLO Lagrangian has the following building blocks
\begin{itemize}
\item one of the ${\cal O}(p^2)$ structures given in (\ref{eq:op2struc});
\item one $\bar T_\mu$ and one $T_\nu$, either contracted with each other or with one of the ${\cal O}_2^{\mu\nu}$ structures 
from (\ref{eq:op2struc});
\item gamma matrices and/or chiral derivatives acting on $T_\nu$ provided that they are contracted with the ${\cal O}_2^{\mu\nu}$ 
structures from (\ref{eq:op2struc});
\item a $\gamma_5$ if required by parity symmetry.
\end{itemize}
We will now go through the various ${\cal O}(p^2)$ structures one by one. 

The $\chi_-$ structure involves a $\gamma_5$; 
see also \cite{Fettes:1998ud} for the corresponding spin-1/2 baryon case. 
However, 
if $\gamma_5$ is the only spinor matrix the resulting expression is further suppressed. 
This can be most easily seen in the Pauli-Dirac 
representation of the Dirac spinors and spinor matrices \cite{bjorken-drell}. Here the lower two components of the Dirac spinors
are small (in the low-energy regime) while $\gamma_5$ is off-diag\-onal and therefore mixes large and small components. 

The $\chi_+$ constitutes a flavor nonet structure, i.e.\ a singlet and an octet. Both couple in a unique way to the 
baryon decuplet and baryon anti-decuplet. Thus we obtain terms $\sim \bar T_{abc}^\mu \, (\hat \chi_+)^c_d \, T_\mu^{abd}$ and 
$\sim \bar T_{abc}^\mu \, T_\mu^{abc} \, {\rm tr}(\chi_+)$. Here 
\begin{eqnarray}
  \label{eq:defchihat}
  \hat\chi_+ := \chi_+ - \frac13 \, {\rm tr}(\chi_+)
\end{eqnarray}
denotes the traceless/octet part of $\chi_+$ \cite{Fettes:1998ud}. 
These terms lead to mass splitting and mass shift, respectively. 
To be in line with the mass term in the LO Lagrangian (\ref{eq:baryonlagr}) we use instead
\begin{eqnarray}
  && -d_{\chi,(8)} \, \bar T_{abc}^\mu \, (\hat \chi_+)^c_d \, \gamma_{\mu\nu} \, (T^\nu)^{abd}   \nonumber \\
  && {} -d_{\chi,(1)} \,  \bar T_{abc}^\mu \, \gamma_{\mu\nu} \, (T^\nu)^{abc} \, {\rm tr}(\chi_+)  \,.
  \label{eq:chi+dec8}
\end{eqnarray}

Next we turn to $f_+^{\mu\nu}$. In principle one finds the following four structures
\begin{eqnarray}
  i \, (\bar T_\mu)_{abc} \, (f^{\mu\nu}_+)^c_d \, T_\nu^{abd} &\,, &
  i \, \bar T_{abc}^\mu \, (f^{\alpha\beta}_+)^c_d \, \gamma_{\alpha\beta}\, T_\mu^{abd} \,, \nonumber \\
  i \, \bar T_{abc}^\mu \, (f^{\alpha\beta}_+)^c_d \, D_\alpha \, D_\beta \, T_\mu^{abd}  & \,, &
  \bar T_{abc}^\mu \, (f^{\alpha\beta}_+)^c_d \, \gamma_\alpha \, D_\beta \, T_\mu^{abd} \,. \phantom{m}
  \label{eq:fplusall}
\end{eqnarray}
In the following we will suppress the flavor structure because it is always the same.
We will now show that two of the terms in (\ref{eq:fplusall}) are suppressed and one is redundant. 
On account of (\ref{eq:supprDD}), (\ref{eq:supprDD2}) the 
term with the two chiral derivatives acting on $T$ is not of NLO. It is of order ${\cal O}(p^4)$, not ${\cal O}(p^2)$.
With the same rewriting as in (\ref{eq:groD}), one can relate the last two terms of (\ref{eq:fplusall}). 
Finally, we will rewrite the second term into the first. To this end we introduce $\tilde f_{\kappa\lambda}^+$ via 
\begin{eqnarray}
  \label{eq:deftildef}
  f^{\alpha\beta}_+ = \epsilon^{\alpha\beta\kappa\lambda} \, \tilde f_{\kappa\lambda}^+ 
\end{eqnarray}
and use \cite{schouten}
\begin{equation}
  \label{eq:schouten-ident-T}
  \bar T^\mu \, \epsilon^{\alpha\beta\kappa\lambda} = 
  \bar T^\alpha \, \epsilon^{\mu\beta\kappa\lambda} + \bar T^\beta \, \epsilon^{\alpha\mu\kappa\lambda} +
  \bar T^\kappa \, \epsilon^{\alpha\beta\mu\lambda} + \bar T^\lambda \, \epsilon^{\alpha\beta\kappa\mu}  \,.
\end{equation}
Using the first constraint equation of (\ref{eq:LO-EOMs}) one finds after some simple rewriting:
\begin{eqnarray}
  \label{eq:sigmafplus}
  i \, \bar T^\mu \, f^{\alpha\beta}_+ \, \gamma_{\alpha\beta}\, T_\mu & = &
  2i \, \bar T_\mu \, \epsilon^{\mu\nu\kappa\lambda} \, \tilde f_{\kappa\lambda}^+ \, T_\nu + {\cal O}(p^3)  \nonumber \\
  & = & 2i \, \bar T_\mu \, f_+^{\mu\nu} \, T_\nu + {\cal O}(p^3)  \,.
\end{eqnarray}
Thus we use as the single $f_+$ structure in a minimal NLO Lagrangian the quantity
\begin{eqnarray}
  \label{eq:NLOmagdec}
  d_M \, i \, (\bar T_\mu)_{abc} \, (f^{\mu\nu}_+)^c_d \, T_\nu^{abd}  \,.
\end{eqnarray}

Next we study the structures $f_-^{\mu\nu}$ and $D^\mu u^\nu + D^\nu u^\mu$ which enforce the presence of a $\gamma_5$. If the 
vector-spinors of the baryons are contracted with these structures, then there are no other gamma matrices and therefore the 
$\gamma_5$ causes a suppression of the terms. 
This leaves us with the terms where the baryon vector-spinors are contracted with each other. Then the structure 
$f_-^{\mu\nu}$ or $D^\mu u^\nu + D^\nu u^\mu$ is (a) contracted with two derivatives or (b) with one gamma matrix and one derivative 
or (c) with two gamma matrices. For case (a) we have the same suppression effect from the $\gamma_5$. Case (b) is forbidden by
charge conjugation symmetry. For case (c) the 
symmetric structure $D^\mu u^\nu + D^\nu u^\mu$ turns the gamma matrices into a metric tensor. Again the $\gamma_5$ suppresses the
term. The antisymmetric structure $f_-^{\mu\nu}$ turns the gamma matrices into $\gamma_{\mu\nu}$. This term is antisymmetric 
with respect to charge conjugation and therefore forbidden.

Finally, we discuss the terms containing $u_\mu u_\nu$. Disregarding for a moment the flavor composition, the independent structures 
are
\begin{eqnarray}
  \label{eq:listdecuu}
  \bar T^\mu u^\alpha u_\alpha T_\mu \,, \quad \bar T^\mu u^\alpha u^\beta \{D_\alpha,D_\beta\} T_\mu
  \,, \quad \bar T^\mu u_\mu u^\nu T_\nu  \,.
\end{eqnarray}
Two more structures are conceivable, but they are redundant: $\bar T^\mu u^\alpha u^\beta \gamma_{\alpha\beta} T_\mu$ 
can be related to the last term in (\ref{eq:listdecuu}) using the same manipulations as in (\ref{eq:sigmafplus}). 
$\bar T^\mu u^\alpha u^\beta \, i \gamma_\alpha D_\beta T_\mu$ can be related to the second term in (\ref{eq:listdecuu})
using the procedure outlined in (\ref{eq:groD}). Note that we have symmetrized the hard derivatives in the second term of 
(\ref{eq:listdecuu}) since the antisymmetrized part is suppressed according to (\ref{eq:supprDD}), (\ref{eq:supprDD2}).

Finally we have to explore the possible flavor structures; see also the discussion in Appendix \ref{sec:sgt}. 
To this end we construct 27-plets and octets from the baryon fields:
\begin{eqnarray}
  && ( \tilde {\cal S}^{\mu\nu} )^{ab}_{a'b'} :=  \bar T^\mu_{a'b'c} \, (T^\nu)^{abc} \nonumber \\ && {} \phantom{mm}
  - \frac15 \, 
  (\delta^{a}_{a'} \, \delta^{e'}_{b'} \, \delta^{b}_{e} + \delta^{a}_{b'} \, \delta^{e'}_{a'} \, \delta^{b}_{e} + 
  \delta^{b}_{a'} \, \delta^{e'}_{b'} \, \delta^{a}_{e} + \delta^{b}_{b'} \, \delta^{e'}_{a'} \, \delta^{a}_{e}  )  \nonumber \\ && 
  \phantom{mmmmm} \times
  \bar T^\mu_{e'dc} \, (T^\nu)^{edc}  \nonumber \\ && {} \phantom{mm}
  + \frac{1}{20} \, (\delta^{a}_{a'} \, \delta^{b}_{b'} + \delta^{a}_{b'} \, \delta^{b}_{a'}) \, \bar T^\mu_{edc} \, (T^\nu)^{edc}  \,,
  \label{eq:27-plet-barT-T}  
\end{eqnarray}
\begin{eqnarray}
  && ({\cal S}'_{\alpha\beta} )^{ab}_{a'b'} :=  \bar T^\mu_{a'b'c} \, (\{D_\alpha,D_\beta\} T_\mu)^{abc} \nonumber \\ && {} \phantom{mm}
  - \frac15 \, 
  (\delta^{a}_{a'} \, \delta^{e'}_{b'} \, \delta^{b}_{e} + \delta^{a}_{b'} \, \delta^{e'}_{a'} \, \delta^{b}_{e} + 
  \delta^{b}_{a'} \, \delta^{e'}_{b'} \, \delta^{a}_{e} + \delta^{b}_{b'} \, \delta^{e'}_{a'} \, \delta^{a}_{e}  )  \nonumber \\ && 
  \phantom{mmmmm} \times
  \bar T^\mu_{e'dc} \, (\{D_\alpha,D_\beta\} T_\mu)^{edc}  \nonumber \\ && {} \phantom{mm}
  + \frac{1}{20} \, (\delta^{a}_{a'} \, \delta^{b}_{b'} + \delta^{a}_{b'} \, \delta^{b}_{a'}) \, 
  \bar T^\mu_{edc} \, (\{D_\alpha,D_\beta\} T_\mu)^{edc}  \nonumber \\ && {} \phantom{mm} + \mbox{terms with $D$'s acting on $\bar T$,}
  \label{eq:27-plet-barT-TDD}  
\end{eqnarray}
\begin{eqnarray}
  \label{eq:octet-barT-T}
  (\tilde {\cal O}^{\mu\nu})^a_{a'} := \bar T^\mu_{a'bc} \, (T^\nu)^{abc} - \frac13 \, \delta^{a}_{a'} \, \bar T^\mu_{dbc} \, (T^\nu)^{dbc}
  \,,
\end{eqnarray}
\begin{eqnarray}
  ({\cal O}'_{\alpha\beta})^a_{a'} & := & \bar T^\mu_{a'bc} \, (\{D_\alpha,D_\beta\} T_\mu)^{abc}   \nonumber \\  && {}
  - \frac13 \, \delta^{a}_{a'} \, \bar T^\mu_{dbc} \, (\{D_\alpha,D_\beta\} T_\mu)^{dbc}  
  \nonumber \\  && {}
  + \mbox{terms with $D$'s acting on $\bar T$.}
  \label{eq:octet-barT-TDD}
\end{eqnarray}

For the meson part we consider the decomposition
of a pair of meson octets into irreducible representations: $8 \otimes 8 = 1_S \oplus 8_S \oplus 8_A \oplus 10_A \oplus \bar{10}_A \oplus 27_S$ 
where $S$ denotes symmetric and $A$ antisymmetric under exchange of the two octet fields. 
In the first two terms of (\ref{eq:listdecuu}), the two meson fields are symmetric under exchange. Thus we only need the 
singlet, the octet ${\cal O}_D$ from (\ref{eq:octetMF}) and the 27-plet from (\ref{eq:27M}). 
This leads to the terms
\begin{eqnarray}
  && d_{1,(1)} \, \bar T^\mu_{abc} \, T_\mu^{abc} \; {\rm tr}(u^\alpha u_\alpha)
  \nonumber \\
  && {} 
  + d_{1,(8)} \, (\tilde {\cal O}^{\mu\nu})^a_{a'} \, g_{\mu\nu} \, ({\cal O}_D^{\alpha\beta})^{a'}_a \, g_{\alpha\beta}  \nonumber \\
  && {}   + d_{1,(27)} \, (\tilde {\cal S}^{\mu\nu})^{ab}_{a'b'} \, g_{\mu\nu} \, ({\cal S}_M^{\alpha\beta})^{a'b'}_{ab} \, g_{\alpha\beta}  
  \nonumber \\
  && {} + d_{2,(1)} \, {\rm tr}(u^\alpha u^\beta) \nonumber \\ && \phantom{m} \times
  \left(\bar T^\mu_{abc} \, (\{D_\alpha,D_\beta\} T_\mu)^{abc} + (\{D_\alpha,D_\beta\} \bar T^\mu)_{abc} \, T_\mu^{abc} \right)
  \nonumber \\
  && {} + d_{2,(8)} \, ({\cal O}'_{\alpha\beta})^a_{a'} \, ({\cal O}_D^{\alpha\beta})^{a'}_a 
  \nonumber \\
  && {} 
  + d_{2,(27)} \, ({\cal S}'_{\alpha\beta})^{ab}_{a'b'} \, ({\cal S}_M^{\alpha\beta})^{a'b'}_{ab} \,.
  \label{eq:uuTT12}
\end{eqnarray}

For the last term in (\ref{eq:listdecuu}), one option would be to build irreducible representations from a baryon and a meson 
field. On account of $8 \otimes 10 = 8 \oplus 10 \oplus 27 \oplus 35$ this yields four independent structures. Yet to be in line with 
the previous construction principles we will build separate irreducible representations from the meson fields and from the 
baryon fields. To this end we use $10 \otimes \bar{10} = 1\oplus8\oplus27\oplus64$, which leads to the four combinations
\begin{eqnarray}
  && d_{3,(1)} \, \bar T^\mu_{abc} \, T_\nu^{abc} \; {\rm tr}(u_\mu u^\nu)
  + d_{3,D} \, (\tilde {\cal O}_{\mu\nu})^a_{a'} \, ({\cal O}_D^{\mu\nu})^{a'}_a   \nonumber \\
  && {} + d_{3,F} \, (\tilde {\cal O}_{\mu\nu})^a_{a'} \, ({\cal O}_F^{\mu\nu})^{a'}_a  
  + d_{3,(27)} \, (\tilde {\cal S}_{\mu\nu})^{ab}_{a'b'} \, ({\cal S}_M^{\mu\nu})^{a'b'}_{ab}  \,. \phantom{mm}   
  \label{eq:uuTT3}  
\end{eqnarray}

A minimal and complete NLO Lagrangian for the pure decuplet sector is given by the sum of the terms in (\ref{eq:chi+dec8}), 
(\ref{eq:NLOmagdec}), (\ref{eq:uuTT12}), and (\ref{eq:uuTT3}).

\subsection{Octet sector}
\label{sec:octetNLO}

The construction of the NLO Lagrangian for baryon-octet chiral perturbation theory has been performed 
in \cite{Oller:2006yh,Frink:2006hx}. With a slight change in labeling the low-energy constants it can be written as
\begin{eqnarray}
  {\cal L}_{8}^{(2)} &=& b_{\chi,D} \, {\rm tr}(\bar{B} \{ \hat\chi_+,B\})  +
  b_{\chi,F} \, {\rm tr}(\bar{B}[ \hat\chi_+,B ]) 
  \nonumber \\
  && {}+  b_{\chi,(1)} \, {\rm tr}(\bar{B} B) \, {\rm tr}(\chi_+)  \nonumber \\
  && {}+
  b_{1,1} \, {\rm tr}(\bar{B} [ u^\mu,[u_\mu,B]])  +
  b_{1,2} \, {\rm tr}(\bar{B}\{ u^\mu,\{u_\mu,B\}\})  \nonumber \\
  && {}+
  b_{1,3} \, {\rm tr}(\bar{B}\{ u^\mu,[u_\mu,B]\})  +
  b_{1,4} \, {\rm tr}(\bar{B}B) \, {\rm tr}( u^\mu u_\mu)  \nonumber \\
  && {}+
  i b_{2,1} \, \big({\rm tr}(\bar{B}[u^\mu,[u^\nu,\gamma_\mu
    {D}_\nu B]])   \nonumber \\
  && \phantom{mmmm} {}-
    {\rm tr}( \bar{B}\overleftarrow{D}_\nu[u^\nu,[u^\mu,\gamma_\mu B]]) \big)  \nonumber \\
  && {}+
  i b_{2,2} \,
  \big({\rm tr}(\bar{B}[u^\mu,\{u^\nu,\gamma_\mu{D}_\nu B\}])  \nonumber \\
  && \phantom{mmmm} {}-
    {\rm tr}( \bar{B}\overleftarrow{D}_\nu\{u^\nu,[u^\mu,\gamma_\mu B]\}) \big)  \nonumber \\
  && {}+
  i b_{2,3} \, \big({\rm tr}(\bar{B}\{u^\mu,\{u^\nu,\gamma_\mu{D}_\nu B\}\})  \nonumber \\
  && \phantom{mmmm} {}-
    {\rm tr}( \bar{B}\overleftarrow{D}_\nu \{u^\nu,\{u^\mu,\gamma_\mu B\}\}) \big)  \nonumber \\
  && {}+
  i b_{2,4} \, \big({\rm tr}(\bar{B}\gamma_\mu{D}_\nu B) -
    {\rm tr}( \bar{B}\overleftarrow{D}_\nu\gamma_\mu B) \big) \, {\rm tr}( u^\mu u^\nu)  \nonumber \\
  && {}+
  \frac{i}{2}\, b_{3,1} \, {\rm tr}(\bar{B} u^\mu) \,  {\rm tr}( u^\nu \sigma_{\mu\nu} B)    \nonumber \\
  && {}+
  \frac{i}{2} \, b_{3,2} \, {\rm tr}(\bar{B}\{[u^\mu,u^\nu],\sigma_{\mu\nu} B\})  \nonumber \\
  && {}+
  \frac{i}{2} \, b_{3,3} \, {\rm tr}(\bar{B}[[u^\mu,u^\nu],\sigma_{\mu\nu} B])     \nonumber \\
  && {}+
  b_{M,D} \, {\rm tr}(\bar{B}\{ f_+^{\mu\nu},\sigma_{\mu\nu} B\})
  \nonumber \\
  && {}+  b_{M,F} \, {\rm tr}(\bar{B}[ f_+^{\mu\nu},\sigma_{\mu\nu} B])   \,.
  \label{eq:NLO1}
\end{eqnarray}
We have nothing to add to this Lagrangian. We just want to compare it to the Lagrangian of the decuplet sector and interpret
its structures from the point of view of irreducible representations. 

The three terms $\sim b_{\chi,\ldots}$ are easy to explain. The octet plus singlet structure of $\chi_+$ is matched by the singlet 
and two octets formed from $\bar B \otimes B$. The corresponding structures for the decuplet sector can 
be found in (\ref{eq:chi+dec8}). The same construction principles concern the (magnetic) field strength terms $\sim b_{M,\ldots}$.
On account of (\ref{eq:sigmafplus}), the corresponding structure (\ref{eq:NLOmagdec}) in the decuplet sector looks 
somewhat different, but is equivalent. 

Turning to the meson-baryon scattering terms we have chosen the labels for the low-energy constants such that terms 
$\sim b_{i,\ldots}$ correspond to the terms $\sim d_{i,\ldots}$ from (\ref{eq:uuTT12}) and (\ref{eq:uuTT3}) for $i=1,2,3$. 
What remains to be understood is the multiplet structure of the $b_{i,\ldots}$ terms. Instead of constructing higher irreducible
representations it is common practice in the octet sector to just multiply $3\times 3$ matrices in all conceivable independent
ways. In practice, the Cayley-Hamilton theorem is very useful to get rid of redundant 
structures \cite{Gasser:1984gg,Bijnens:1999sh,Ebertshauser:2001nj}. 
Yet one should find with both methods the same number of terms --- though in different linear combinations. This is what we will
show now.

The terms $\sim b_{1,\ldots}$ and $\sim b_{2,\ldots}$ are essentially symmetric when exchanging the meson fields. 
For the $b_{2,\ldots}$ terms this can be better seen when rewriting $i \gamma^\mu D^\nu$ terms to $\{D^\mu, D^\nu\}$; see the 
corresponding discussion in subsection \ref{sec:decNLO} after (\ref{eq:listdecuu}). Thus the mesons can build the irreducible 
representations of a singlet, octet, and 27-plet. The baryon fields can build one singlet, two octets, and one 27-plet. Thus one 
finds 4 independent structures --- in line with the four terms $\sim b_{1,\ldots}$ and the four terms $\sim b_{2,\ldots}$. 

In the terms $\sim b_{3,\ldots}$ the $\sigma_{\mu\nu}$ antisymmetrizes the meson fields. The corresponding irreducible 
representations are an octet, a decuplet, and an anti-decuplet. The baryon fields provide two octets, one decuplet, and one 
anti-de\-cu\-plet. The octet structures can be seen in the terms $\sim b_{3,2}$ and $\sim b_{3,3}$. The combination of a meson 
decuplet and a baryon anti-decuplet is neither hermitian nor symmetric with respect to charge conjugation symmetry. Thus one 
has to build one proper combination of the decuplets and anti-decuplets. In total, this yields a third $b_{3,\ldots}$ term.

Thus we have found that the method of building irreducible representations yields a consistent number of NLO terms for 
the baryon-octet sector.

\section{Determination of some low-energy constants}
\label{sec:detLEC}

In this and the next section we will determine some of the low-energy constants, i.e.\ some of the parameters 
of the NLO Lagrangian. 

We start with the flavor-breaking terms $\sim \chi_+$ as given in (\ref{eq:chi+dec8}) and (\ref{eq:NLO1}).
At leading order, all members of one bayon multiplet have the same mass. The flavor breaking terms provide mass splitting 
and overall mass shifts for both multiplets. 

If all meson and external fields are put to zero, then $\chi_+$ reduces to $4 B_0 {\cal M}$ 
with the quark mass matrix ${\cal M}$. In the 
following we ignore isospin breaking. Typically isospin breaking effects are in size comparable to electromagnetic corrections.
Thus it would not be reasonable to consider one while ignoring the other. In the isospin limit, one finds 
\begin{eqnarray}
  \label{eq:qmmat}
  {\cal M} = \left(
    \begin{array}{ccc}
      m_q & 0 & 0 \\ 0 & m_q & 0 \\ 0 & 0 & m_s
    \end{array}
  \right)
\end{eqnarray}
and \cite{Gasser:1984gg}
\begin{equation}
  \label{eq:goldstone-masses}
  m_\pi^2 = 2 m_q B_0 + {\cal O}(p^4) \,, \quad m_K^2 = (m_q+m_s) B_0 + {\cal O}(p^4) 
\end{equation}
with the (isospin averaged) mass of the pion/kaon, $m_{\pi/K}$.

\begin{table*}
  \centering
  \begin{tabular}[t]{|c|llllllllll|}
    \hline
    \phantom{$\int^O_P$} & $\Delta^{++}$ & $\Delta^+$ & $\Delta^0$ & $\Delta^-$ & $\Sigma^{*+}$ & $\Sigma^{*0}$ & $\Sigma^{*-}$ & 
    $\Xi^{*0}$ & $\Xi^{*-}$ & $\Omega$
    \\     \hline
    NLO & $4.44$ & $2.22$ & $0$ & $-2.22$ & $2.14$ & $0$ & $-2.14$ & $0$ & $-2.07$ & $-2.02$ \\
    exp. & $3.7$ to $7.5$ & $2.7 \pm 3.5$ & --- & --- & --- & --- & --- & --- & --- & $-2.02 \pm 0.05$  \\ \hline         
  \end{tabular}
  \caption{NLO predictions and experimental values \cite{pdg} for the magnetic moments of the decuplet baryons in units of 
    nuclear magnetons. The magnetic moment of the $\Omega$ baryon is fitted.}
  \label{tab:mdm-dec}
\end{table*}
\begin{table*}
  \centering
  \begin{tabular}[t]
{|c|lllllllll|}
    \hline
    \phantom{$\int^O_P$} & $p$ & $n$ & $\Lambda$ & $\Sigma^+$ & $\Sigma^0$ & $\Sigma^-$ & $\Xi^0$ & $\Xi^-$ & $\Sigma^0 \Lambda$ 
    \\     \hline
    NLO & $2.75$ & $-1.61$ & $-0.80$ & $2.54$ & $0.80$ & $-0.92$ & $-1.61$ & $-0.85$ & $1.39$ \\
    exp. & $2.793(0)$ & $-1.913(0)$ & $-0.613(4)$ & $2.458(10)$ & --- & $-1.160(25)$ & $-1.250(14)$ & $-0.651(3)$ & $\pm1.61(8)$  \\ \hline     
  \end{tabular}
  \caption{NLO fit and experimental values \cite{pdg} for the magnetic moments of the octet baryons in units of 
    nuclear magnetons. The last column provides the magnetic transition moment for $\Sigma^0 \to \Lambda$.}
  \label{tab:octetMDM}
\end{table*}
At NLO accuracy one, finds the following relations between the decuplet masses and the NLO parameters
\begin{eqnarray}
  \label{eq:massesNLOd}
  && d_{\chi,(8)} = \frac38 \, \frac{m_\Omega - m_{\Sigma^*}}{m_K^2-m_\pi^2}  \,, \\
  \label{eq:massesNLOSigma}
  && m_{\Sigma^*} = m_{(10)} + (2m_\pi^2 + 4 m_K^2) \, d_{\chi,(1)}  \,, \\
  \label{eq:massesNLODelta}
  && m_\Delta = m_{\Sigma^*} - \frac12 \, (m_\Omega - m_{\Sigma^*})  \,, \\
  && m_{\Xi^*} = m_{\Sigma^*} + \frac12 \, (m_\Omega - m_{\Sigma^*})  \,.
  \label{eq:massesNLOXi}
\end{eqnarray}
Note that $\chi_+$ produces also meson-baryon four-point interactions. Thus, in principle, the mass-shift parameter $d_{\chi,(1)}$ 
can be determined from meson-baryon scattering phase shifts. In practice, however, there are no direct data on meson-baryon 
scattering for the spin-3/2 baryons.

Numerically we obtain $d_{\chi,(8)} \approx 0.47 \,$GeV$^{-1}$. This constitutes a quite natural value 
considering the typical hard scale of about 
$1\,$GeV as set by the nucleon mass or by the scale of chiral symmetry breaking, $4\pi F_\pi$. 
For the right-hand side of (\ref{eq:massesNLODelta}) and 
(\ref{eq:massesNLOXi}) we find $1.24 \,$GeV and $1.53 \,$GeV, respectively, in very reasonable agreement with the experimental 
values for the masses of $\Delta$ and $\Xi^*$ \cite{pdg}. 

The same quality can be obtained in the octet sector. Again, the term $\sim b_{\chi,(1)}$ from (\ref{eq:NLO1}) 
leads to an overall mass shift, while the octet terms $\sim b_{\chi,D}$ and $\sim b_{\chi,F}$ provide the mass splitting, 
in line with \cite{GellMann:1961ky,okubo-mass,GellMann:1964xy}. 
Numerical values for the shifted mass and for the splitting parameters are 
given, for instance, in \cite{Kubis:2000aa}: 
\begin{eqnarray}
  && b_{\chi,D} \approx  0.060  \, {\rm GeV}^{-1}  \,, \\
  && b_{\chi,F} \approx  -0.190  \, {\rm GeV}^{-1} \,,  \\
  && m_{(8)} - (2m_\pi^2+4m_K^2) \, b_{\chi,(1)}  \approx  1.192 \, {\rm GeV}.
  \label{eq:valKubismasses}
\end{eqnarray}

The term (\ref{eq:NLOmagdec}) provides an anomalous magnetic moment for the decuplet states. It is worth to point out why 
this is at NLO a magnetic interaction and not an electric one. As can be most easily seen in the 
Pauli-Dirac representation \cite{bjorken-drell} the constraint equations of (\ref{eq:LO-EOMs}) suppress the $T^0$ component 
relative to the spatial components of the vector-spinor field. Thus the NLO part of (\ref{eq:NLOmagdec}) contains only the 
field $f_+^{ij}$, which is related to the magnetic field strength on account of (\ref{eq:emvA}). The same is true for the 
terms $\sim b_{M,D/F}$ in (\ref{eq:NLO1}). They provide (flavor symmetric) anomalous magnetic moments for the octet baryons, 
in line with the seminal calculations of \cite{Coleman:1961jn}. The decuplet-to-octet transitions 
from (\ref{eq:transNLO}) feature a magnetic transition moment $\sim c_M$ and an axial vector electric transition moment 
$\sim c_E$. 

We stress again that at NLO there is only one term, $\sim d_M$, that provides an anomalous magnetic moment for the 
decuplet states,
one term, $\sim c_M$, that provides a magnetic transition moment, and one term, $\sim c_E$, that provides an 
axial vector electric transition moment. 
Sometimes in the literature, more terms have been written down, but at NLO accuracy they are all degenerate with one of 
these three terms. If one carries out an NNLO calculation it is more transparent to write down the pertinent NNLO interaction
terms instead of carrying along NNLO differences between redundant NLO terms.

Concerning magnetic moments, one-loop calculations constitute the present state of the 
art \cite{Geng:2009ys,Kubis:2000aa,Geng:2008mf,Geng:2009hh}. Of course, these calculations contain also the NLO tree-level
terms that emerge from the Lagrangian of the present work. One can fit the NLO calculation to data or a calculation that 
includes one-loop corrections (NNLO or N$^{3}$LO). Of course, the results obtained for the numerical values of the 
NLO parameters will be different. To keep the present NLO work self-consistent we provide here only the values for the NLO 
parameters based on an NLO calculation. For improved calculations including loop effects we refer 
to \cite{Geng:2009ys,Kubis:2000aa,Geng:2008mf,Geng:2009hh} and references therein. 

In the decuplet sector the best known magnetic moment comes from the $\Omega$ baryon \cite{pdg}. Following essentially 
\cite{Geng:2009ys} we fit $d_M$ to the magnetic moment of the $\Omega$ and provide results for the other 
magnetic moments. Given the fact that our NLO Lagrangian provides a proper mass splitting for the decuplet states 
we use physical masses throughout. Therefore our results do not fit exactly with those results of \cite{Geng:2009ys} that are 
called ``SU(3) symmetric''. Introducing the magnetic moment $\mu_D$ normalized to the nuclear 
magneton $\mu_N:=e/(2m_N)$, $m_N \approx 0.94\,$GeV,
\begin{eqnarray}
  \label{eq:decmdm-norm}
  \hat\mu_D := \frac{\mu_D}{\mu_N}  \,,
\end{eqnarray}
we obtain the following NLO relation between the normalized magnetic moments
\begin{eqnarray}
  \label{eq:mdm-rel-dec}
  \hat\mu_D = q_D \, \left( \frac{m_N}{m_D}-\frac{m_N}{m_\Omega}-\hat\mu_\Omega \right)  
\end{eqnarray}
where $q_D$ denotes the electric charge of the decuplet state $D$ and $m_D$ its mass. 
Results are provided in table \ref{tab:mdm-dec}.
The NLO low-energy constant is given by
\begin{eqnarray}
  \label{eq:dMdetOmega}
  d_M = \frac{3}{4m_N} \left(\hat\mu_\Omega + \frac{m_N}{m_\Omega} \right) \approx -1.16 \,\mbox{GeV}^{-1} \,.
\end{eqnarray}

In the octet sector, we fit $\sim b_{M,D/F}$ to the measured magnetic 
moments \cite{pdg}. Note that there is a subtle difference to the results of \cite{Kubis:2000aa,Geng:2008mf} called 
${\cal O}(q^2)$ or ``tree level'' therein. In \cite{Kubis:2000aa,Geng:2008mf} one overall baryon-octet mass is used.
Based on the fact that our NLO Lagrangian provides the proper mass splitting for the octet baryons 
we use physical masses throughout. In other words, the anomalous magnetic moments are flavor symmetric, but the contribution 
provided by the respective charge is properly weighted by the mass of the respective baryon. We find
\begin{eqnarray}
  \label{eq:valbMs}
  b_{M,D} \approx 0.321 \, \mbox{GeV}^{-1} \,, \quad b_{M,F} \approx 0.125 \, \mbox{GeV}^{-1}
\end{eqnarray}
and the magnetic moments given in table \ref{tab:octetMDM}.
We observe agreement on a level of $\pm$30\%. 
Of course, the description cannot be perfect at NLO level. See also \cite{Kubis:2000aa,Geng:2008mf} for discussions about the 
importance and size of loop effects.

The magnetic transition moments are discussed in the next section. Finally, we note that the axial form factors for the transition
of a $\Delta$ to a nucleon are addressed in \cite{Geng:2008bm} at the one-loop level. 
In principle, this calculation involves the low-energy 
constant $c_E$ of (\ref{eq:transNLO}). In lack of corresponding high-quality data, it is difficult to pin down $c_E$. We note,
however, that the $c_E$ term also gives rise to a four-point interaction that couples a photon and a meson to the baryons. Thus 
it enters, for instance, the decay processes $B(J=3/2) \to B'(J=1/2) \pi \gamma$. Those processes receive LO and NLO 
contributions, which allows for some uncertainty estimates even in the absence of data for such decays. Measuring such decays, 
for instance in the hyperon sector, will help to provide estimates for $c_E$ and other parameters of the decuplet sector. 
Calculating the three-body decay distributions at NLO accuracy is, however, beyond the scope of the present work.

\section{Radiative transitions}
\label{sec:predictrad}

At NLO all the radiative decays $B(J=3/2) \to B'(J=1/2) \, \gamma$ are given by just one interaction 
term $\sim c_M$ according to the corresponding Lagrangian (\ref{eq:transNLO}) and the replacement (\ref{eq:emvA}). 
Thus one can determine $c_M$ by comparing to the measured decay widths for 
$\Delta \to N \, \gamma$, $\Sigma^{*+} \to \Sigma^+ \, \gamma$ and $\Sigma^{*0} \to \Lambda \, \gamma$ \cite{pdg}. 

The partial decay width of a decuplet baryon with mass $M$ decaying into an octet baryon with mass $m$ and a photon
is given by
\begin{equation}
  \Gamma = \frac{c^2}{6\pi} p_{\text{c.m.}}^3 \frac{E_{B'} + M}{M} \,,
  \label{eq:raddec}
\end{equation}
where $E_{B'} = \sqrt{m^2 + p_{\text{c.m.}}^2}$  ($p_{\text{c.m.}}$) is the energy (momentum) of the outgoing baryon in the rest frame of 
the decaying resonance. For the calculation, one needs the properties of the spin-3/2 vector-spinors collected in 
Appendix \ref{sec:vecspin}. 

At NLO, the coefficient $c$ that appears in \eqref{eq:raddec} is given by $c_M e$ times a flavor factor. For the various 
radiative decays these flavor factors are provided in table \ref{tab:radTrans}.
\begin{table}[H]
  \centering
  \small
  {\tabulinesep=0.5mm
    \begin{tabu} {|l|rrr|}
      \hline
      Decay & $c/(c_M e)$ & BR [\%] & $c_M$ [GeV$^{-1}$] \\ \hline
      $\Delta \to N\gamma$	  & $2/{\sqrt{3}}$ & 0.60$\pm$0.05 & 2.00$\pm$0.03\\
      $\Sigma^{*+} \to \Sigma^+\gamma$	& $-2/\sqrt{3}$ & 0.70$\pm$0.17 & 1.89$\pm$0.08\\
      $\Sigma^{*-} \to \Sigma^-\gamma$	& $0$  &  $<0.024$ & --- \\
      $\Sigma^{*0} \to \Sigma^0\gamma$	& $1/\sqrt{3}$ & \textbf{0.18$\pm$0.01} & ---  \\
      $\Sigma^{*0} \to \Lambda\gamma$	 & $-1$  & 1.25$\pm$0.13 & 1.89$\pm$0.05\\
      $\Xi^{*0} \to \Xi^0\gamma$ & $-2/\sqrt{3}$ & \textbf{4.0$\pm$0.3} & --- \\
      $\Xi^{*-} \to \Xi^-\gamma$ & $0$ & $<4$ & --- \\ \hline
    \end{tabu}}
  \caption{NLO predictions (in bold) and experimental values \cite{pdg} for the branching ratios of the radiative decuplet 
    decays (third column). 
    The second column shows the calculated flavor factors and the last column contains $c_M$ as determined from the 
    respective experimental value.}
  \label{tab:radTrans}
\end{table}
Matching to the measured decay widths yields the average $c_M = (1.92 \pm 0.08)\,$GeV$^{-1}$. 
With this input, one can make an NLO prediction for the unknown decays widths $\Sigma^{*0} \to \Sigma^0 \gamma$ and 
$\Xi^{*0} \to \Xi^0 \gamma$. This is provided in table \ref{tab:radTrans} as the bold entries. 

Unfortunately, at NLO one cannot obtain a non-trivial prediction for the decays $\Sigma^{*-} \to \Sigma^- \, \gamma$ and 
$\Xi^{*-} \to \Xi^- \, \gamma$. These decays break flavor symmetry, or more specifically U-spin symmetry 
(see also, e.g., \cite{Kaxiras:1985zv}). Since in \eqref{eq:transNLO} the NLO interaction term $\sim c_M$ is completely 
flavor symmetric, U-spin symmetry is exact.

Let us give some details on the role of U-spin for the considered decays: Like isospin concerns up and down quarks, U-spin 
concerns down and strange quarks. The four negatively charged decuplet states $\Delta^-$, $\Sigma^{*-}$, $\Xi^{*-}$, and $\Omega$ 
form a U-spin quartet. The two negatively charged octet states $\Sigma^-$ and $\Xi^-$ constitute a U-spin doublet. But the photon
is a U-spin singlet since down and strange quark has the same electric charge. U-spin symmetry forbids the transition from 
a quartet to a doublet plus singlet. 

Of course, U-spin symmetry is broken at some point in the chiral expansion. For the radiative decays this happens at 
NNLO. For the one-loop contributions to NNLO the mesons in the loops carry different masses as obtained from the mesonic 
LO chiral Lagrangian \cite{Gasser:1984gg,Scherer:2002tk,Scherer:2012xha}. This will lead to non-vanishing decay widths for the 
radiative decays $\Sigma^{*-} \to \Sigma^- \, \gamma$ and $\Xi^{*-} \to \Xi^- \, \gamma$. Interestingly, one can expect that 
the results from the loop calculation will be finite, i.e.\ do not require renormalization. The reason is that flavor-symmetry 
breaking counter terms for these processes appear only at N$^3$LO since such structures must involve a symmetry breaking 
term $\chi_\pm$ and a field strength $f^{\mu\nu}_+$, which in total is at least of order ${\cal O}(p^4)$.
Such calculations are beyond the scope of the present NLO work.
Without being able to give numerical values for the decays widths $\Sigma^{*-} \to \Sigma^- \, \gamma$ 
and $\Xi^{*-} \to \Xi^- \, \gamma$, one can at least conclude that these widths are small compared to the predictions for the 
other radiative decays.

\begin{acknowledgement}
{\bf Acknowledgements:} SL thanks H.\ Ghaderi for very valuable discussions on group theory.
\end{acknowledgement}

\appendix

\section{Some group theory}
\label{sec:sgt}

We want to find the proper flavor structures for the third term of (\ref{eq:gen3}). 
The first step is to combine the two mesons to irreducible representations \cite{Georgi-groupth,greiner-mueller} 
and to do the same for the two baryons. One obtains
$8 \otimes 8 = 1 \oplus 8  \oplus 8\oplus 10 \oplus \bar{10} \oplus 27$ and $8 \otimes 10 = 8 \oplus 10 \oplus 27 \oplus 35$. Thus to construct a singlet for an 
interaction Lagrangian one has four possibilities: First, to take 27-plets from the mesons and the baryons. 
Second, to take the anti-decuplet from the mesons and the decuplet from the baryons. Finally, to take one from the two-meson octets and the octet from the baryons. 

The general properties of the relevant multiplets are: The octet has one upper and one lower index and is traceless.
(One might re-interpret the upper and lower index as row and column index, respectively.) 
We recall that upper indices transform with $h$ under flavor transformations while the lower
components transform with $h^\dagger$. 

The (anti-)decuplet has three upper (lower) indices and is fully symmetric in all indices. Indeed this 
yields ten states: 3 states with all indices identical, 6 states with two indices identical, 1 state with all three indices 
different. $3+6+1=10$.

The 27-plet has two upper and two lower indices. It is symmetric in its upper indices and also symmetric in its lower ones. 
In addition, it is traceless with respect to any contraction of an upper with a lower index. In fact, these are 27 independent
states: One has 6 combinations for the upper and 6 for the lower components. The fact that the tensor should be traceless 
places 9 constraints (the various choices for the two non-contracted indices). $6*6-9=27$.

For completeness, we note that the 35-plet has four upper and one lower index. It is fully symmetric in its upper indices 
and traceless with respect to any contraction of an upper with the lower index. The condition that the trace vanishes takes 
away 10 degrees of freedom. Thus one should find 45 combinations. The lower index allows for 3 choices. Thus one should 
get 15 combinations from having four fully symmetrized upper indices. Indeed, there are 3 states with all indices the same, 
6 states with three indices the same, 3 states with two indices the same and 
also the other two the same, and 3 states with two indices the same and the others mutually different and to the pair. 
$3*(3+6+3+3)-10=35$.

We start with the octet constructed from the baryons: 
\begin{eqnarray}
  \label{eq:octetB}
  ({\cal O}_{\mu\nu})^a_d = \varepsilon_{deb} \, \bar B^e_c  \, \gamma_\mu \gamma_5 \, T^{abc}_\nu \,.
\end{eqnarray}
The corresponding decuplet is
\begin{equation}
  \label{eq:decuB}
  {\cal D}^{abd}_{\mu\nu} = \bar B^d_c  \, \gamma_\mu \gamma_5 \, T^{abc}_\nu 
  + \bar B^a_c  \, \gamma_\mu \gamma_5 \, T^{bdc}_\nu 
  + \bar B^b_c  \, \gamma_\mu \gamma_5 \, T^{dac}_\nu  \,.
\end{equation}
The 27-plet is given by
\begin{eqnarray}
  && ({\cal S}_{\mu\nu})^{ab}_{de} = 
  (\varepsilon_{eic} \, \bar B^i_d + \varepsilon_{dic} \, \bar B^i_e)  \, \gamma_\mu \gamma_5 \, T^{abc}_\nu \nonumber \\ && {}
  - \frac15 \, \bar B^i_j  \, \gamma_\mu \gamma_5 \, 
  (\varepsilon_{eic} \, \delta^a_d \, T^{jbc}_\nu + \varepsilon_{dic} \, \delta^a_e \, T^{jbc}_\nu + 
  \varepsilon_{eic} \, \delta^b_d \, T^{jac}_\nu    \nonumber \\ && \phantom{mmmmmmm} {} 
  + \varepsilon_{dic} \, \delta^b_e \, T^{jac}_\nu)  \,. 
  \phantom{mm} 
  \label{eq:27B}
\end{eqnarray}

The octets constructed from the meson fields $(u^\mu)$ and $(u^\nu)$ can be chosen to be symmetric and antisymmetric,
respectively, with respect to an exchange of the two fields:
\begin{eqnarray}
  \label{eq:octetMF}
  ({\cal O}_F^{\mu\nu})^a_b &=& (u^\mu)^a_j \, (u^\nu)^j_b - (u^\nu)^a_j \, (u^\mu)^j_b   \,, \\
  ({\cal O}_D^{\mu\nu})^a_b &=& (u^\mu)^a_j \, (u^\nu)^j_b + (u^\nu)^a_j \, (u^\mu)^j_b 
  - \frac23 \, \delta^a_b \, (u^\mu)^k_j \, (u^\nu)^j_k  \,. \nonumber 
\end{eqnarray}
The mesonic anti-decuplet is
\begin{eqnarray}
  (\bar{\cal D}_M^{\mu\nu})_{abc} &=& \varepsilon_{aij} \, (u^\mu)^i_b \, (u^\nu)^j_c + \varepsilon_{bij} \, (u^\mu)^i_c \, (u^\nu)^j_a 
  \nonumber \\ && {}
  + \varepsilon_{cij} \, (u^\mu)^i_a \, (u^\nu)^j_b 
  + \varepsilon_{aij} \, (u^\mu)^i_c \, (u^\nu)^j_b 
  \nonumber \\ && {}
  + \varepsilon_{bij} \, (u^\mu)^i_a \, (u^\nu)^j_c + \varepsilon_{cij} \, (u^\mu)^i_b \, (u^\nu)^j_a  \,. 
  \label{eq:adecuM}
\end{eqnarray}
Finally the mesonic 27-plet is given by
\begin{eqnarray}
  && ({\cal S}_M^{\mu\nu})^{ab}_{cd} = (u^\mu)^a_c \, (u^\nu)^b_d + (u^\mu)^b_c \, (u^\nu)^a_d \nonumber \\ && {}
  + (u^\mu)^a_d \, (u^\nu)^b_c + (u^\mu)^b_d \, (u^\nu)^a_c  \nonumber \\ && {}
  - \frac15 \, \delta^a_c \, \Big((u^\mu)^b_j \, (u^\nu)^j_d + (u^\mu)^j_d \, (u^\nu)^b_j\Big) \nonumber \\ && {}
  - \frac15 \, \delta^a_d \, \Big((u^\mu)^b_j \, (u^\nu)^j_c + (u^\mu)^j_c \, (u^\nu)^b_j\Big) \nonumber \\ && {} 
  - \frac15 \, \delta^b_c \, \Big((u^\mu)^a_j \, (u^\nu)^j_d + (u^\mu)^j_d \, (u^\nu)^a_j\Big) \nonumber \\ && {}
  - \frac15 \, \delta^b_d \, \Big((u^\mu)^a_j \, (u^\nu)^j_c + (u^\mu)^j_c \, (u^\nu)^a_j\Big) \nonumber \\ && {} 
  + \frac{1}{10} \, (\delta^a_c \, \delta^b_d + \delta^b_c \, \delta^a_d) \,  (u^\mu)^j_k \, (u^\nu)^k_j  \,.
  \label{eq:27M}  
\end{eqnarray}

\section{Discrete symmetries}
\label{sec:discsym}

In the following, we list how fermion bilinears transform with respect to parity $P$ and charge conjugation $C$. 
We denote the parity transformation matrix for Lorentz vectors by $({\cal P}^{\mu\nu})$. It is defined 
like the metric tensor $(g^{\mu\nu})$, but with an additional sign flip for the spatial components. 
Note that $T^\mu$ transforms with an extra minus relative to $B$ concerning parity, 
but not concerning charge conjugation. 

First, we study bilinears built from a spin-1/2 antiparticle field $\bar B$ and 
a spin-3/2 particle field $T^\mu$ \cite{Rarita:1941mf,deJong:1992wm,Hacker:2005fh,Pascalutsa:2006up}. 
We find
\begin{eqnarray}
  \bar B \, (\gamma_5) \, T^\mu & \stackrel{P}{\to} & \mp {\cal P}^{\mu \mu'} \, \bar B \, (\gamma_5) \, T_{\mu'}  \,, \\
  \bar B \, \gamma^\nu (\gamma_5) \, T^\mu & \stackrel{P}{\to} & 
  \mp {\cal P}^{\mu \mu'} \, {\cal P}^{\nu \nu'} \, \bar B \, \gamma_{\nu'} (\gamma_5) \, T_{\mu'}  \,, \\
  \bar B \, \sigma^{\alpha\beta} (\gamma_5) \, T^\mu & \stackrel{P}{\to} & 
  \mp {\cal P}^{\mu \mu'} \, {\cal P}^{\alpha \alpha'} \, {\cal P}^{\beta \beta'} \, 
  \bar B \, \sigma_{\alpha'\beta'} (\gamma_5) \, T_{\mu'}  
  \label{eq:listP}
\end{eqnarray}
and 
\begin{eqnarray}
  \bar B^a_b \, (\gamma_5) \, (T^\mu)^{cde} & \stackrel{C}{\to} & \bar T^\mu_{cde} \, (\gamma_5) \, B^b_a  \,, \\
  \bar B^a_b \, \gamma^\nu (\gamma_5) \, (T^\mu)^{cde} & \stackrel{C}{\to} & 
  \mp \bar T^\mu_{cde} \, \gamma^{\nu} (\gamma_5) \, B^b_a  \,, \\
  \bar B^a_b \, \sigma^{\alpha\beta} (\gamma_5) \, (T^\mu)^{cde} & \stackrel{C}{\to} & 
  -\bar T^\mu_{cde} \, \sigma^{\alpha\beta} (\gamma_5) \, B^b_a \,. 
  \label{eq:listC}
\end{eqnarray}
The upper/lower sign refers to the case without/with $\gamma_5$. 

Next we turn to the bilinears built from a pair of spin-3/2 fields. These combinations transform as
\begin{eqnarray}
  \bar T^\mu \, (\gamma_5) \, T^\nu & \stackrel{P}{\to} & 
  \pm {\cal P}^{\mu \mu'} \, {\cal P}^{\nu \nu'} \, \bar T_{\mu'} \, (\gamma_5) \, T_{\nu'}  \,, \\
  \bar T^\mu \, \gamma^\alpha (\gamma_5) \, T^\nu & \stackrel{P}{\to} & 
  \pm {\cal P}^{\mu \mu'} \, {\cal P}^{\alpha \alpha'} \, {\cal P}^{\nu \nu'} \, 
  \bar T_{\mu'} \, \gamma_{\alpha'} (\gamma_5) \, T_{\nu'}  \,, \\
  \bar T^\mu \, \sigma^{\alpha\beta} (\gamma_5) \, T^\nu & \stackrel{P}{\to} & 
  \pm {\cal P}^{\mu \mu'} \, {\cal P}^{\alpha \alpha'} \, {\cal P}^{\beta \beta'} \, {\cal P}^{\nu \nu'} \, 
  \bar T_{\mu'} \, \sigma_{\alpha'\beta'} (\gamma_5) \, T_{\nu'}  \nonumber \\ &&  
  \label{eq:listP2}
\end{eqnarray}
and 
\begin{eqnarray}
  \bar T^\mu_{abc} \, (\gamma_5) \, (T^\nu)^{def} & \stackrel{C}{\to} & \bar T^\nu_{def} \, (\gamma_5) \, (T^\mu)^{abc}  \,, \\
  \bar T^\mu_{abc} \, \gamma^\nu (\gamma_5) \, (T^\nu)^{def} & \stackrel{C}{\to} & 
  \mp \bar T^\nu_{def} \, \gamma^{\nu} (\gamma_5) \, (T^\mu)^{abc}  \,, \\
  \bar T^\mu_{abc} \, \sigma^{\alpha\beta} (\gamma_5) \, (T^\nu)^{def} & \stackrel{C}{\to} & 
  -\bar T^\nu_{def} \, \sigma^{\alpha\beta} (\gamma_5) \, (T^\mu)^{abc} \,. 
  \label{eq:listC2}
\end{eqnarray}
As before, the upper/lower sign refers to the case without/with $\gamma_5$.

\section{Vector-spinors}
\label{sec:vecspin}

Spin-3/2 states are described by vector-spinors \cite{Rarita:1941mf,deJong:1992wm,Hacker:2005fh,Pascalutsa:2006up}.
They satisfy 
\begin{eqnarray}
  \label{eq:complspin32}
  \sum\limits_\sigma u_\mu(p,\sigma) \, \bar u_\nu(p,\sigma) = - (\slashed{p}+m) \, P^{3/2}_{\mu\nu}(p)
\end{eqnarray}
where $p^0=\sqrt{m^2+\vec p^2}$ denotes the energy of the particle described by the vector-spinor and $m$ its mass. 
The projector on spin 3/2 is defined by
\begin{eqnarray}
  \label{eq:defproj32aa}
  P^{3/2}_{\mu\nu}(p) := g_{\mu\nu} - \frac13 \, \gamma_\mu \gamma_\nu 
    - \frac{1}{3 p^2} \, (\slashed{p} \, \gamma_\mu \, p_\nu + p_\mu \, \gamma_\nu \, \slashed{p})  \,.
\end{eqnarray}
Note that for (\ref{eq:complspin32}), the scalar product $p^2$ appearing in (\ref{eq:defproj32aa}) can be replaced by $m^2$.

\bibliography{lit}{}
\bibliographystyle{epj}
\end{document}